\documentclass[a4paper,aps,prb,nobibnotes,superscriptaddress,twocolumn]{revtex4}
\usepackage{amsmath,amsbsy,amsfonts,mathrsfs,color,graphicx,epsfig}

\newcommand{\D}{\mathrm{d}}

\renewcommand{\Im}{\mathfrak{Im}\,}

\newcommand{\mint}[1]{\int\!\! \D^{3} #1 \, }

\newcommand{\ba}{\begin{array}}
\newcommand{\ea}{\end{array}}

\begin{document}

\title{On the use of Neumann's principle for the calculation of the polarizability tensor of nanostructures}

\author{Micael\,J.\,T. Oliveira}
\thanks{Author to whom correspondence should be addressed. E-mail: micael@teor.fis.uc.pt}
\affiliation{European Theoretical Spectroscopy Facility (ETSF)}
\affiliation{Centre of Computational Physics, University of Coimbra, 
Rua Larga, 3004-516 Coimbra Portugal}
\affiliation{Departamento de F\'{\i}sica de Materiales,
Facultad de Ciencias Qu\'{\i}micas (UPV/EHU), Centro Mixto CSIC-UPV/EHU and
Donostia International Physics Center, San Sebasti\'an 20018, Spain}

\author{Alberto Castro}
\affiliation{European Theoretical Spectroscopy Facility (ETSF)}
\affiliation{Institut f\"ur Theoretische Physik, Freie Universit\"at Berlin,
  Arnimallee 14, 14195 Berlin, Germany}

\author{Miguel\,A.\,L. Marques}
\affiliation{European Theoretical Spectroscopy Facility (ETSF)}
\affiliation{Centre of Computational Physics, University of Coimbra, 
Rua Larga, 3004-516 Coimbra Portugal}

\author{Angel Rubio}
\affiliation{European Theoretical Spectroscopy Facility (ETSF)}
\affiliation{Departamento de F\'{\i}sica de Materiales,
Facultad de Ciencias Qu\'{\i}micas (UPV/EHU), Centro Mixto CSIC-UPV/EHU and
Donostia International Physics Center, San Sebasti\'an 20018, Spain}

\begin{abstract}
The polarizability measures how the system responds to an applied
electrical field. Computationally, there are many different ways to
evaluate this tensorial quantity, some of which rely on the explicit
use of the external perturbation and require several individual
calculations to obtain the full tensor. In this work, we present some
considerations about symmetry that allow us to take full advantage of
Neumann's principle and decrease the number of calculations required
by these methods. We illustrate the approach with two examples, the
use of the symmetries in real space and in spin space in the
calculation of the electrical or the spin response.
\end{abstract}

\maketitle

\section{Introduction}

The redistribution of electrons in a finite system that occurs when it
is exposed to an external electromagnetic field is characterized by a
set of constants called
polarizabilities\cite{polarizability-book}. The {\em static}
polarizabilities refer to static fields, whereas the {\em dynamical}
polarizabilities are functions of the frequency of the external
field. Usually the name {\em polarizability} is restricted to the
constants that determine the cited redistribution to first order in
the applied field -- when high-intensity fields are applied, one needs
to make use of the higher-order ones or {\em hyperpolarizabilities}.

Besides characterizing the electromagnetic field-matter interaction,
they are also important when studying collision phenomena, and as
coarse indicators of physical size, structure and shape. The knowledge
of the polarizability helps to obtain numerous physical quantities
that depend on it: the dielectric constant and the refractive index of
extended systems, the long-range interaction potentials between
polarizable systems, van der Waals constants, etc.  The dynamical
polarizability, also, contains precious information about the excited
states of a finite system: the peaks of the polarizability as a
function of the energy determine the excitation energies of the
system.  The absorption cross section of a finite system is trivially
related to the imaginary part of the dynamical
polarizability\cite{polarizability-book,TDDFT}. The oscillator
strengths of the transition are the weights of the peaks of this
absorption spectrum.

The polarizabilities may also be classified as dipole, quadrupole,
etc., depending on the shape of the perturbing field: For example, the
dipole polarizabilities characterize the response of the system to a
dipole field. Rigorously speaking, one should also specify the
physical magnitude that is measured, and speak, for example, of the
dipole-dipole polarizability: the response of the system dipole to a
dipole field.  In this Article, we will exclusively deal with
dipole-dipole first-order polarizabilities -- the most commonly
studied ones, and arguably the most important for nanostructure
characterisation.  Since there are three possible directions for the
perturbing dipole field, and three possible directions for the system
dipole response, the dipole-dipole polarizability is in fact a
three-dimensional tensor.

Because of its tensorial character, one way of simplifying the
calculation of the polarizability is to use Neumann's
principle~\cite{neumann}: the polarizability tensor of the system must
be left invariant under any transformation that is also a point
symmetry operation of the system.  This condition of invariance
reduces the number of independent tensor components, since it
signifies relationships between those components, thus potentially
reducing the number of calculations necessary to obtain the full
tensor.

Numerous theoretical techniques can be used to calculate
polarizabilities, with varying level of accuracy and detail. In
particular, there is a class of methods that rely on the explicit use
of the external perturbation, i.e. each line of the tensor is obtained
by performing one calculation. Because of this, when using these
methods it is not always obvious along which directions the perturbing
fields should be aplied in order to make full use of Neumann's
principle. In this Article we discuss how this can be done.

{\em Stricto sensu}, the polarizabilities refer to the reaction to
spin-independent (i.e., {\em electrical}) perturbations measured by
spin-independent observables; they are referred to as {\em
density-density} response functions.  However, one can also think of
more general response functions and apply spin-dependent perturbations
and/or look at spin-dependent observables, obtaining in this way {\em
spin-density}, {\em density-spin} and {\em spin-spin} response
functions -- these generalized objects are sometimes referred to as
{\em susceptibilities}.  Even though this work is mainly concerned
with the polarizability itself, we will consider these more general
objects, since we will also show how the calculation of the {\em
density-density} and {\em spin-spin} response of a spin-saturated
molecular system may be simplified.

In the following section, we recall the necessary definitions, and
list some of the first-principles techniques that can be used to
calculate polarizabilities of technologicaly relevant nano and bio
structures -- some of which can benefit from the simplifications
proposed in this Article. These are discussed in
Section~\ref{section:method}. Section~\ref{section:spin} shows how
essentially the sames ideas may serve to simplify the calculation of
the singlet and triplet excitations of systems whose ground-state is
spin-saturated. Finally, we present two systems where the technique
was used to compute the density-density and the spin-spin responses.

\section{The polarizability tensor}
\label{section:definitions}

We will now introduce the notation and key quantities that are
relevant for the use of symmetries to obtain the different linear
response functions. Let $n_{\sigma}\;(\sigma = \uparrow,\downarrow)$
be the spin-density of a system of $N$ electrons:
\begin{equation}
  n_{\sigma}(\vec{r},t) = \langle \Phi(t)\vert   
  \sum_{k=1}^N \delta(\vec{r}-\hat{\vec{r}}_k)\delta_{\sigma\sigma_k}
  \vert\Phi(t)\rangle\,,
\end{equation}
If we apply an infinitesimal perturbation, $\delta
v_{\sigma}(\vec{r},\omega)$ ($\sigma=\uparrow,\downarrow$), the
response of the density, $\delta n_{\sigma}(\vec{r}, t) =
n_{\sigma}(\vec{r},t)-n_{\sigma}(\vec{r},0)$, will be related to the
perturbation, to first order, by a density response function,
$\chi$. In the frequency domain this is expressed as:
\begin{equation}
  \label{eq:density_response}
  \delta n_{\sigma}(\vec{r}, \omega) = \sum_{\sigma'}
  \mint{r'} \chi_{\sigma\sigma'}(\vec{r},\vec{r}',\omega) \delta v_{\sigma'}(\vec{r}',\omega)\,.
\end{equation}
The variation of the {\em total density}, $n =
n_\uparrow+n_\downarrow$, and of the {\em magnetization density}, $m =
n_\uparrow-n_\downarrow$, are given by:
\begin{subequations}
  \begin{align}
    \delta n(\vec{r}, \omega) & = \delta n_\uparrow(\vec{r},\omega) + \delta n_\downarrow(\vec{r}, \omega)\,,
    \\
    \delta m(\vec{r}, \omega) & = \delta n_\uparrow(\vec{r},\omega) - \delta n_\downarrow(\vec{r}, \omega)\,.
  \end{align}
\end{subequations}

After the system is perturbed, one can obtain information about the
system by looking at the variation of some observable: $\delta
\langle\hat{\mathcal{O}}\rangle(t) =
\langle\hat{\mathcal{O}}\rangle(t) -
\langle\hat{\mathcal{O}}\rangle(0)$.  In our case, we will be looking
at the dipole of the system in each of the spatial directions,
$\hat{X}_i$. In order to learn about the spin modes, one may
also look at $\hat{X}_i\hat{\sigma}_z$, where $\hat{\sigma}_z$ is the
Pauli $z$-matrix. In the frequency domain, the behavior of these
observables is given by:
\begin{subequations}
  \begin{align}
    \label{eq:dipole_response}
    \delta\langle\hat{X}_i\rangle(\omega) & =
    \mint{r} x_i \delta n(\vec{r},\omega)\,.
    \\
    \label{eq:spin_dipole_response}
    \delta\langle\hat{X}_i\hat{\sigma}_z\rangle(\omega) & =
    \mint{r} x_i \delta m(\vec{r},\omega)\,.
  \end{align}
\end{subequations}

One also has to define which kind of perturbations are to be
considered: We will restrict hereafter the formulation to dipole
perturbations of two kinds: \renewcommand{\labelenumi}{(\roman{enumi})}

\noindent (i)~{\em Spin-independent perturbations} of the form:
\begin{equation}
  \label{eq:spin_independent}
  \delta v^{[n]}_{\sigma}(\vec{r},\omega) =  - x_j \kappa(\omega)\,.
\end{equation}
In this case, the variations $\delta n$ and $\delta m$ are:
\begin{subequations}
  \begin{align}
    \delta n(\vec{r},\omega) & = 
    - \kappa(\omega) 
    \mint{r'} \chi^{[nn]}(\vec{r},\vec{r}',\omega) x'_j\,,
    \\
    \delta m(\vec{r},\omega) & = 
    - \kappa(\omega) 
    \mint{r'} \chi^{[mn]}(\vec{r},\vec{r}',\omega) x'_j\,,
  \end{align}
\end{subequations}
where we have defined the new objects:
\begin{subequations}
  \begin{align}
    \chi^{[nn]} & = \chi_{\uparrow\uparrow} + \chi_{\uparrow\downarrow} 
    + \chi_{\downarrow\uparrow} + \chi_{\downarrow\downarrow}
    \\
    \chi^{[mn]} & = \chi_{\uparrow\uparrow} + \chi_{\uparrow\downarrow} 
    - \chi_{\downarrow\uparrow} - \chi_{\downarrow\downarrow}
  \end{align}
\end{subequations}

The observables $\delta\langle\hat{X}_i\rangle(\omega)$ and
$\delta\langle\hat{X}_i\hat{\sigma}_z\rangle(\omega)$ will be given
by:
\begin{subequations}
  \begin{align}
    \delta\langle\hat{X}_i\rangle^{[n]}_j(\omega) & = - \kappa(\omega)
    \int\!\!\!\int\!\! {\rm d}^3r\,{\rm d}^3r'\, x_i
    \chi^{[nn]}(\vec{r},\vec{r}',\omega) x'_j\,,
    \\
    \delta\langle\hat{X}_i\hat{\sigma}_z\rangle^{[n]}_j(\omega) & =
    -\kappa(\omega) \int\!\!\!\int\!\! {\rm d}^3r\,{\rm d}^3r'\, x_i
    \chi^{[mn]}(\vec{r},\vec{r}',\omega) x'_j\,.
  \end{align}
\end{subequations}
The superscript ``${[n]}$'' means that the observable is measured
after a spin-independent perturbation of the form given in
Eq.~\eqref{eq:spin_independent} has been applied, whereas the
subscript ``$j$'' means that this perturbation has been applied in the
direction $j$.

\noindent(ii)~{\em Spin-dependent perturbations} of the form:
\begin{equation}
  \label{eq:spin_dependent}
  \delta v^{[m]}_{\sigma}(\vec{r},\omega) = 
  \left\{
  \begin{array}{r@{\quad,\quad}l}
    -x_j\kappa(\omega) & \sigma = \uparrow
    \\
    x_j\kappa(\omega) & \sigma = \downarrow\,.
  \end{array}
  \right.
\end{equation}
Or, making use of the Pauli $z$-matrix:
\begin{equation}
  \delta v^{[m]}(\vec{r},\omega) = -x_j\kappa(\omega)\sigma_z\,.
\end{equation}
The variations of $n(\vec{r},\omega)$ and $m(\vec{r},\omega)$ are:
\begin{subequations}
  \begin{align}
    \delta n(\vec{r},\omega) & =
    - \kappa(\omega) 
    \mint{r'} \chi^{[nm]}(\vec{r},\vec{r}',\omega) x'_j\,,
    \\
    \delta m(\vec{r},\omega) & =
    - \kappa(\omega) 
    \mint{r'} \chi^{[mm]}(\vec{r},\vec{r}',\omega) x'_j\,,
  \end{align}
\end{subequations}
with the definitions:
\begin{subequations}
  \begin{align}
    \chi^{[nm]} & = \chi_{\uparrow\uparrow} - \chi_{\uparrow\downarrow} 
    + \chi_{\downarrow\uparrow} - \chi_{\downarrow\downarrow}
    \\
    \chi^{[mm]} & = \chi_{\uparrow\uparrow} - \chi_{\uparrow\downarrow} 
    - \chi_{\downarrow\uparrow} + \chi_{\downarrow\downarrow}
  \end{align}
\end{subequations}
Now, the observables $\delta\langle\hat{X}_i\rangle(\omega)$ and
$\delta\langle\hat{X}_i\hat{\sigma}_z\rangle(\omega)$ will be given
by
\begin{subequations}
  \begin{align}
    \delta\langle\hat{X}_i\rangle^{[m]}_j(\omega) & =  - \kappa(\omega) \int\!\!\!\int\!\! {\rm d}^3r{\rm d}^3r
    x_i \chi^{[nm]}(\vec{r},\vec{r}',\omega) x'_j\,,
    \\
    \delta\langle\hat{X}_i\hat{\sigma}_z\rangle^{[m]}_j(\omega) & = -\kappa(\omega) \int\!\!\!\int\!\! {\rm d}^3r{\rm d}^3r
    x_i \chi^{[mm]}(\vec{r},\vec{r}',\omega) x'_j\,.
  \end{align}
\end{subequations}

We now look at the quotients between the induced variations
$\delta\langle\hat{X}_i\rangle(\omega)$ and
$\delta\langle\hat{X}_i\hat{\sigma}_z\rangle(\omega)$ and the strength
of the perturbation, $\kappa(\omega)$, for each one of the cases:
\begin{subequations}
  \begin{align}
    \label{eq:pol_1}
    \alpha_{ij}^{[nn]}(\omega) & = \frac{\delta\langle\hat{X}_i\rangle^{[n]}_j(\omega)}{\kappa(\omega)}
    \\
    \label{eq:pol_2}
    \alpha_{ij}^{[mn]}(\omega) & = \frac{\delta\langle\hat{X}_i\hat{\sigma}_z\rangle^{[n]}_j(\omega)}{\kappa(\omega)}
    \\
    \label{eq:pol_3}
    \alpha_{ij}^{[nm]}(\omega) & = \frac{\delta\langle\hat{X}_i\rangle^{[m]}_j(\omega)}{\kappa(\omega)} 
    \\
    \label{eq:pol_4}
    \alpha_{ij}^{[mm]}(\omega) & = \frac{\delta\langle\hat{X}_i\hat{\sigma}_z\rangle^{[m]}_j(\omega)}{\kappa(\omega)}
  \end{align}
\end{subequations}
The usual definition of the polarizability refers to the first of
these expressions, $\alpha_{ij} \equiv \alpha_{ij}^{[nn]}$. However,
one may also be interested in the other kinds of responses. We will
use the same notation ($\alpha$) for these response functions,
although the name {\em polarizability} should be restricted for the
first case.  Another way of defining these functions is:
\begin{equation}
  \label{eq:polarizability_def}
  \alpha_{ij}^{\sigma\sigma'}(\omega) = 
  - \int\!\!\!\int\!\! {\rm d}^3r\,{\rm d}^3r'\, x_i \chi_{\sigma\sigma'}(\vec{r},\vec{r}',\omega) x'_j\,.
\end{equation}
Obviously, the two definitions are related, and one may retrieve the 
$\alpha_{ij}^{[xy]}$ components from the
$\alpha_{ij}^{\sigma\sigma'} $ and vice\-versa:
\begin{subequations}
  \begin{align}
    \alpha_{ij}^{[nn]} & = \alpha_{ij}^{\uparrow\uparrow} + \alpha_{ij}^{\uparrow\downarrow} +
    \alpha_{ij}^{\downarrow\uparrow} + \alpha_{ij}^{\downarrow\downarrow}
    \\
    \alpha_{ij}^{[mn]} & = \alpha_{ij}^{\uparrow\uparrow} + \alpha_{ij}^{\uparrow\downarrow} - 
    \alpha_{ij}^{\downarrow\uparrow} - \alpha_{ij}^{\downarrow\downarrow}
    \\
    \alpha_{ij}^{[nm]} & = \alpha_{ij}^{\uparrow\uparrow} - \alpha_{ij}^{\uparrow\downarrow} + 
    \alpha_{ij}^{\downarrow\uparrow} - \alpha_{ij}^{\downarrow\downarrow}
    \\
    \alpha_{ij}^{[mm]} & = \alpha_{ij}^{\uparrow\uparrow} - \alpha_{ij}^{\uparrow\downarrow} -
    \alpha_{ij}^{\downarrow\uparrow} + \alpha_{ij}^{\downarrow\downarrow}
  \end{align}
\end{subequations}

Regarding its spatial structure -- and thus dropping the spin indexes
-- the dynamical polarizability elements may be arranged in a
second-rank 3x3 symmetric tensor $\boldsymbol{\alpha}(\omega)$.  The
absorption cross-section tensor is then simply proportional to its
imaginary part:
\begin{equation}
  \boldsymbol{\sigma}(\omega) = \frac{4\pi\omega}{c}\Im \boldsymbol{\alpha}(\omega)\,,
\end{equation}
where $c$ is the speed of light.

The static or the dynamical (hyper)polarizabilities can be calculated
within different theoretical approaches. Our main concern here is
(time-dependent) density functional theory -- (TD)DFT\cite{TDDFT}, but
our arguments are quite general and apply equally well to other
electronic structure methods. The simplest of these is perhaps
obtaining the static (hyper)polarizabilities through finite
differences, i.e., by applying small static electrical fields
$\boldsymbol{E}_j$, and then calculating numerically the
derivatives. On the other hand, the dynamical polarizability can be
calculated through real-time propagation of the time-dependent
Kohn-Sham equations\cite{yabana}. Moreover, both static and dynamical
polarizabilities can be obtained through straighforward perturbation
theory. In this case, ``perturbation theory'' refers to the
application of Sternheimer equation\cite{sternheimer,sternheimer4} in
one way or another, either for the static or for the
dynamical\cite{td-sternheimer,Xavier} case. Another very recent and
quite promising approach, is a efficient Lanczos-based
method\cite{Baroni}.

For all these methods the calculations rely on the explicit use of the
external perturbation. If we want to calculate the full tensor, we
have to perform three calculations, one for each spatial
direction. If, moreover, we need both the density and spin modes, we
have to make two calculations per spatial direction, one for each of
the two possible perturbations discussed above. In the next sections
we will show how the number of actual calculations can be severely
reduced when using these methods by taking advantage of the symmetries
of the system.

Note that the polarizability can also be obtained, for example, from
the response functions, using, for example, the formalism developed by
M.~Casida\cite{casida,casida2,casida3,casida4}. This is a case in
which the considerations presented below will not simplify the
calculations, since they do not proceed by the successive application
of the different perturbations.

\section{Spatial symmetries}
\label{section:method}

In this section, we will drop the spin indices, since the whole
argument is valid for any of the spin components of the
polarizability.

By making use of Eq.~\eqref{eq:polarizability_def}, it is easy to
prove that $\boldsymbol{\alpha}(\omega)$ is a proper tensor: If we
consider a second orthonormal reference frame $\lbrace \hat{e}'_1,
\hat{e}'_2, \hat{e}'_3\rbrace$, $\boldsymbol{\alpha}(\omega)$
transforms following the tensorial transformation law:
\begin{equation}
  \boldsymbol{\alpha}'(\omega) = \mathbf{P}^t\boldsymbol{\alpha}(\omega)\mathbf{P}\,,
\end{equation}
where $\boldsymbol{\alpha}'(\omega)$ is the polarizability in the
second reference frame, and $\mathbf{P}$ is the rotation matrix
between the two frames.

This tensorial character of the polarizability permits us to work in
any orthonormal reference frame; once we obtain its values, we may
easily transform it by straightforward matrix manipulation. We may
then choose the frame which is most appropriate, bearing in mind the
geometry of the molecule, and this can reduce the total number of
calculations.  However, this liberty does not allow us to make full
use of symmetry. For this purpose, we need to work with
non-orthonormal directions.

Let us consider three linearly-independent, but possibly
non-orthogonal, unit vectors $\lbrace
\hat{p}_1,\hat{p}_2,\hat{p}_3\rbrace$.  We define the polarizability
elements $\tilde{\alpha}_{uv}(\omega)$ as:
\begin{equation}
  \tilde{\alpha}_{uv}(\omega) = -\int\!\!\!\int\!\! {\rm d}^3r\,{\rm d}^3r'\, 
  (\vec{r}\cdot\hat{p}_u) \chi(\vec{r},\vec{r}',\omega) 
  (\vec{r}'\cdot\hat{p}_v)\,.
\end{equation}
This corresponds to a process in which the polarization of the
perturbing field is along $\hat{p}_v$, and the dipole is measured
along $\hat{p}_u$. If we know the 3x3 matrix
$\tilde{\boldsymbol{\alpha}}(\omega)$, we get the {\em real} tensor
$\boldsymbol{\alpha}(\omega)$ by making use of the following simple
relationship, which can be obtained once again from
Eq.~\eqref{eq:polarizability_def}:
\begin{equation}
  \label{eq:fundamental_equation}
  \tilde{\boldsymbol{\alpha}}(\omega) =
  \mathbf{P}^t\boldsymbol{\alpha}(\omega)\mathbf{P}\,.
\end{equation}
$\mathbf{P}$ is the transformation matrix between the original
orthonormal reference frame and $\lbrace
\hat{p}_1,\hat{p}_2,\hat{p}_3\rbrace$. Note that this transformation
is in general not a rotation; $\mathbf{P}$ is not unitary. Moreover,
no matter how familiar it looks, Eq.~\eqref{eq:fundamental_equation}
is not a change of coordinates: $\tilde{\boldsymbol{\alpha}}(\omega)$
is {\em not} the polarizability tensor in the new reference frame.
And finally, also note that the traces of
$\tilde{\boldsymbol{\alpha}}$ and $\boldsymbol{\alpha}$ do not
coincide:
\begin{equation}
  {\rm Tr}\left[\tilde{\boldsymbol{\alpha}}(\omega)\right] = 
  {\rm Tr}\left[\mathbf{P}^t\boldsymbol{\alpha}(\omega)\mathbf{P}\right] = 
  {\rm Tr}\left[\boldsymbol{\alpha}(\omega)\mathbf{P}\mathbf{P}^t\right]\,.
\end{equation}
but $\mathbf{P}\mathbf{P}^t \neq \boldsymbol{1}$. Notwithstanding this,
it tells us that we may obtain the polarizability tensor by
calculating the related object $\tilde{\boldsymbol{\alpha}}(\omega)$.

Now let us assume that the molecule under study possesses some
non-trivial symmetry transformations -- to start with, we consider
that it has two, $\mathcal{A}$ and $\mathcal{B}$. We consider an
initial unit vector, $\hat{p}_1$, and define:
\begin{eqnarray}
\nonumber \hat{p}_2 = \mathcal{A}\hat{p}_1\\
          \hat{p}_3 = \mathcal{B}\hat{p}_2
\end{eqnarray}
We assume that this may be done such that the set $\lbrace
\hat{p}_1,\hat{p}_2,\hat{p}_3\rbrace$ is linearly independent.

We then perform a TDDFT calculation with the perturbing field
polarized in the direction $\hat{p}_1$.  This permits us to obtain the
row $\lbrace
\tilde{\alpha}_{11},\tilde{\alpha}_{12},\tilde{\alpha}_{13}\rbrace$.
Since the matrix is symmetric, we also have the column $\lbrace
\tilde{\alpha}_{11},\tilde{\alpha}_{21},\tilde{\alpha}_{31}\rbrace$.
The symmetry of the molecule also permits us to obtain the diagonal:
$\lbrace \tilde{\alpha}_{33} = \tilde{\alpha}_{22} =
\tilde{\alpha}_{11}\rbrace$.  The only missing element is
$\tilde{\alpha}_{23} = \tilde{\alpha}_{32}$, but it is easy to prove
that:
\begin{equation}
  \label{eq:missing_component}
  \tilde{\alpha}_{23} = \tilde{\alpha}_{1,\mathcal{A}^{-1}\hat{p}_3}\,,
\end{equation}
which we can also get from our original calculation. The conclusion
is that we have access to the full tensor by making only one
calculation.

To fix ideas, we use the example of a molecule with one $n$-th order
axis of symmetry ($n>2$). Let $\mathcal{R}$ be the rotation of
$2\pi/n$ degrees around this axis. We then choose $\hat{p}_1$ not
collinear with this axis, and also not perpendicular to it. If we
define $\mathcal{R}\hat{p}_1 = \hat{p}_2$ and $\mathcal{R}\hat{p}_2 =
\hat{p}_3$, the set $\lbrace \hat{p}_1,\hat{p}_2,\hat{p}_3\rbrace$
will be linearly independent.  In this case, moreover, since
$\mathcal{A}=\mathcal{B}=\mathcal{R}$,
Eq.~\eqref{eq:missing_component} reduces to $\tilde{\alpha}_{23} =
\tilde{\alpha}_{12}$.

It may very well be that we may only find two linearly independent
``equivalent axis'', $\hat{p}_1$ and $\hat{p}_2$, related by a
symmetry transformation, $\mathcal{A}$ -- this is the case of a system
that possesses only a plane of symmetry, or only an axis of symmetry
of order two.  We may then define $\hat{p}_1$ to be a tilted vector
with respect to this plane (not contained in it, and not
perpendicular). Then, $\hat{p}_2=\mathcal{A}\hat{p}_1$, where
$\mathcal{A}$ is the reflection on the plane, is an equivalent vector,
and $\vec{p}_3$ can be chosen to lie in the symmetry plane (the
obvious choice will be $\hat{p}_3 = \hat{p}_1 \wedge \hat{p}_2$, that
ensures linear independence).  We then only need two calculations, one
with the polarization along $\hat{p}_1$ (or $\hat{p}_2$) and another
with the polarization along $\hat{p}_3$. Moreover, if $\hat{p}_1$ is
chosen to be tilted exactly $\pi/2$ with respect to the symmetry
plane, the system of vectors is orthonormal, and we do not even need
to apply Eq.~\eqref{eq:fundamental_equation}.  Note that this case
applies to {\em all} planar molecules.

\section{Singlet and triplet excitations in spin-saturated systems}
\label{section:spin}

We consider a system whose ground state is spin-saturated. It
verifies:
\begin{subequations}
  \begin{align}
    \alpha_{ij}^{\uparrow\uparrow} & = \alpha_{ij}^{\downarrow\downarrow}\,,
    \\
    \alpha_{ij}^{\uparrow\downarrow} & = \alpha_{ij}^{\downarrow\uparrow}\,.
  \end{align}
\end{subequations}
And, in consequence, $\alpha_{ij}^{[nm]}=\alpha_{ij}^{[mn]}=0$, and
\begin{subequations}
  \begin{align}
    \alpha_{ij}^{[nn]} & = 2\alpha_{ij}^{\uparrow\uparrow} + 2\alpha_{ij}^{\uparrow\downarrow}\,,
    \\
    \alpha_{ij}^{[mm]} & = 2\alpha_{ij}^{\uparrow\uparrow} - 2\alpha_{ij}^{\uparrow\downarrow}\,.
  \end{align}
\end{subequations}
Despite this symmetry, in order to obtain all spin-components (which
in this spin-saturated case are only two independent ones), by making
use of the two types of perturbations defined in
Eqs.~\eqref{eq:spin_independent} and \eqref{eq:spin_dependent}, we
would still need two calculations: one perturbing with a
spin-independent potential -- in order to obtain $\alpha_{ij}^{[nn]}$,
and one with a spin-dependent one -- in order to obtain
$\alpha_{ij}^{[mm]}$.

\begin{figure}[t]
  \centering
  \epsfig{file=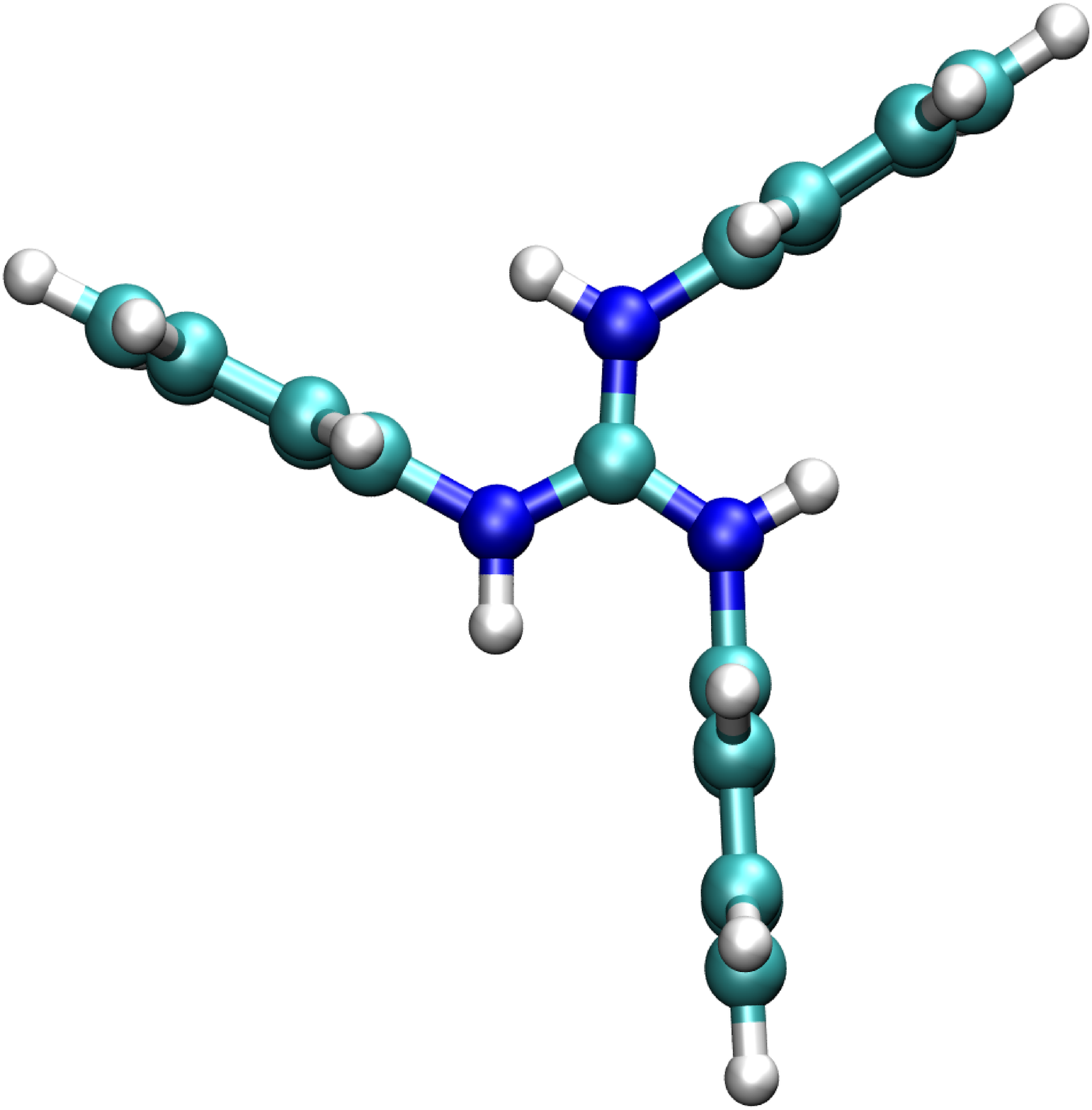,width=0.38\textwidth}

  \vspace{0.5cm}

  \epsfig{file=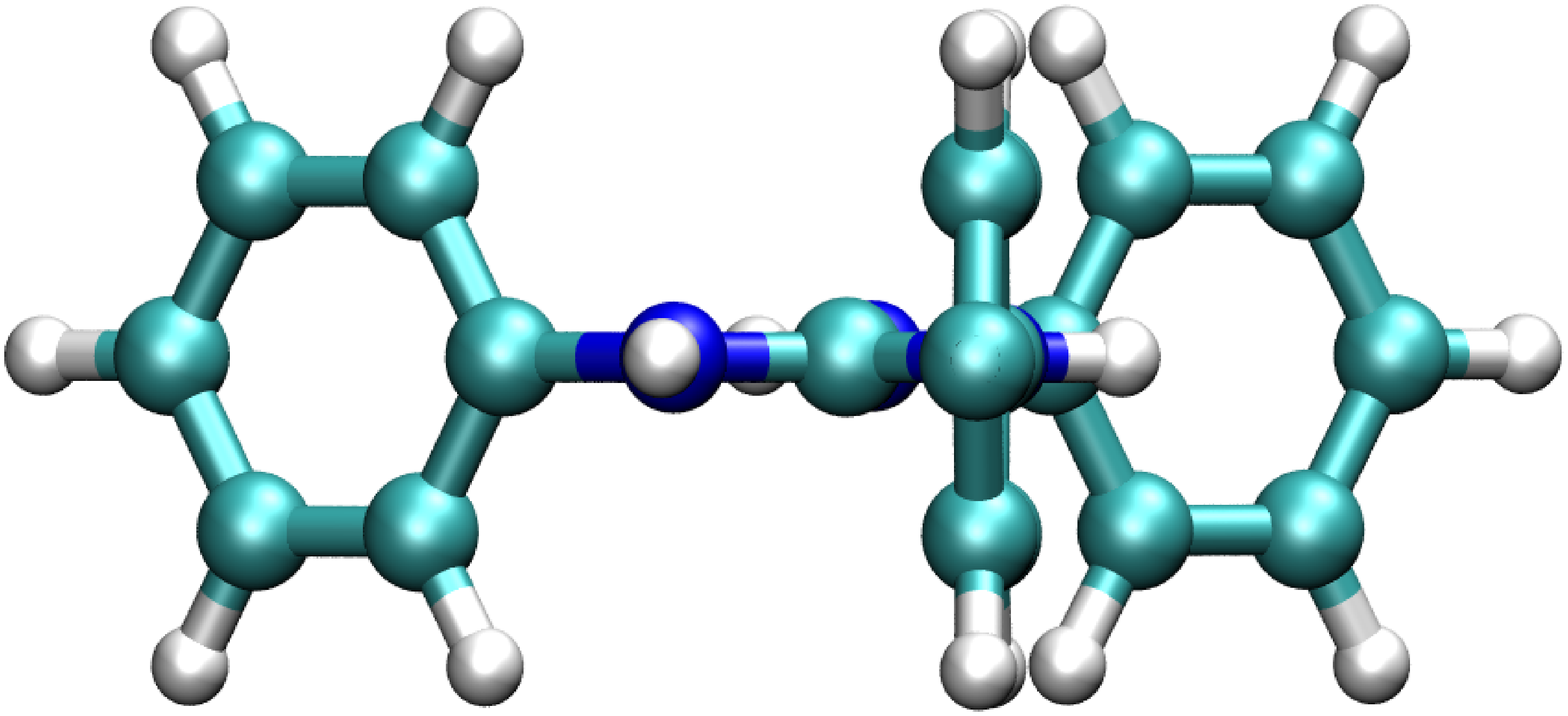,width=0.38\textwidth}
  \caption{(Color online) Two different views of protonated
    triphenylguanidine (point group $C_{3h}$). In this picture, the
    dark blue atoms are nitrogen, the cyan represent carbon, and
    finally the white atoms are hydrogen.
    \label{fig:tpg_geo}}
\end{figure}

However, one can easily use a similar scheme to the one outlined in
the previous section in order to calculate the two polarizabilities in
only one shot. The idea is to apply a perturbation in the form:
\begin{equation}
  \label{eq:spin_perturbation}
  \delta v^{[\uparrow]}_{\sigma}(\vec{r},\omega) = 
  \left\{
  \begin{array}{r@{\quad,\quad}l}
    -x_j\kappa(\omega) & \sigma = \uparrow
    \\
    0 & \sigma = \downarrow\,,
  \end{array}
  \right.
\end{equation}
or, in the Pauli matrix language:
\begin{equation}
  \delta v^{[\uparrow]}(\vec{r},\omega) = -\frac{1}{2}\kappa(\omega)x_j(\hat{\sigma}_0 + \hat{\sigma}_z)\,.
\end{equation}
It is then easy to verify that the response of the dipole observables
will then be given by:
\begin{subequations}
  \begin{align}
    \label{eq:spin-sat_singlets}
    \delta \langle \hat{X}_i \rangle^{[\uparrow]}_j(\omega) = - \frac{1}{2}\kappa(\omega)\alpha_{ij}^{[nn]}\,,
    \\
    \label{eq:spin-sat_triplets}
    \delta \langle \hat{X}_i\hat{\sigma}_z \rangle^{[\uparrow]}_j(\omega) 
    = - \frac{1}{2}\kappa(\omega)\alpha_{ij}^{[mm]}\,,
  \end{align}
\end{subequations}
thus providing us with the components of the two response functions
that we need with only one calculation.

\section{Examples}
\label{section:example}

\begin{figure}[t]
  \centering
  \epsfig{file=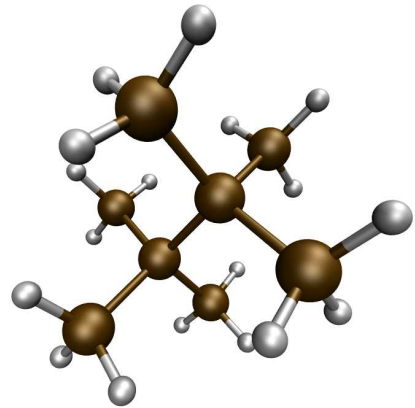,width=0.38\textwidth}

  \caption{(Color online) One view of Si$_8$H$_{18}$ (point group
    $D_{3d}$). In this picture, the dark brown atoms are silicon and
    the grey atoms are hydrogen.
    \label{fig:si8h18_geo}}
\end{figure}

There are many complex molecules of tecnological relevance that
present symmetries such that the schemes outlined in the previous
sections can be used. We chose two of them to illustrated the method:
protonated triphenylguanidine and one hidrogenated silicon cluster
Si$_8$H$_{18}$. Triphenylguanidine compounds are regarded as
interesting for quadratic nonlinear optical applications while
hidrogenated silicon is an important optico-electronic material with
potentially important technological applications.

The ground-state of protonated triphenylguanidine is spin-saturated,
has one proper axis of symmetry of order three, one plane of symmetry,
and one improper axis of rotation of order three (see
Fig.~\ref{fig:tpg_geo}). The ground-state of Si$_8$H$_{18}$ is
spin-saturated, has an inversion center, one proper axis of symmetry
of order three, three proper axis of symmetry of order two, three
planes of symmetry, and one improper axis of symmetry of order six
(see Fig.~\ref{fig:si8h18_geo}). This means one can make use of the
schemes outlined in the previous sections to obtain all the components
of the response functions with only one calculation: all components of
both $\boldsymbol{\alpha}^{[nn]}(\omega)$ and
$\boldsymbol{\alpha}^{[mm]}(\omega)$ tensors.

Without using the symmetry of the system the response functions were
computed by applying spin-independent and spin-dependent perturbations
from Eqs.~\eqref{eq:spin_independent} and \eqref{eq:spin_dependent}
with polarization directions along the $x$, $y$ and $z$
directions. This way the response functions were straighforwardly
obtained but required a total of six time-propagations.

To use the symmetry a set $\lbrace \hat{p}_1, \hat{p}_2, \hat{p}_3
\rbrace$ is needed. In the case of protonated triphenylguanidine we
built it by defining $\hat{p}_1$ to be a vector tilted $\pi/4$ with
respect to the plane of symmetry and the two symmetry transformations
$\mathcal{A}$ and $\mathcal{B}$ to be an inversion with respect to the
plane and a $2\pi/3$ rotation around the axis of symmetry of order 3.
In the case of Si$_8$H$_{18}$ we chose $\hat{p}_1$ to be a vector
tilted $\pi/4$ with respect to the axis of symmetry of order three and
both symmetry transformations $\mathcal{A}$ and $\mathcal{B}$ to be
$2\pi/3$ rotations around the same axis.

Applying a perturbation of the same form as
Eq.~\eqref{eq:spin_perturbation} with a polarization direction along
$\hat{p}_1$ and using Eqs.~\eqref{eq:fundamental_equation},
\eqref{eq:spin-sat_singlets} and \eqref{eq:spin-sat_triplets} allowed
us to obtain the response functions with just one calculation in both
cases.

All response calculations were done with the code {\tt
  octopus}~\cite{octopus} using the Perdew-Zunger~\cite{lda}
parametrization of the adiabatic local density approximation for the
exchange-correlation potential. This method has already been
successfully used for the calculation of optical spectra in a variety
of systems, ranging from small molecules\cite{molecules} and
clusters\cite{c20} to biological systems\cite{bio}.

\begin{figure}[t]
  \centering
  \epsfig{file=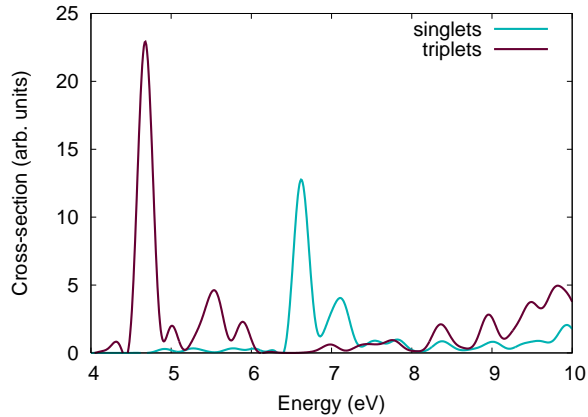,width=0.45\textwidth}
  \caption{(Color online) Protonated triphenylguanidine averaged
    absorption cross-section for singlets and triplets. The curves
    obtained with and without the use of symmetry completely overlap.
    \label{fig:tpg_response}}
\end{figure}

\begin{figure}[t]
  \centering
  \epsfig{file=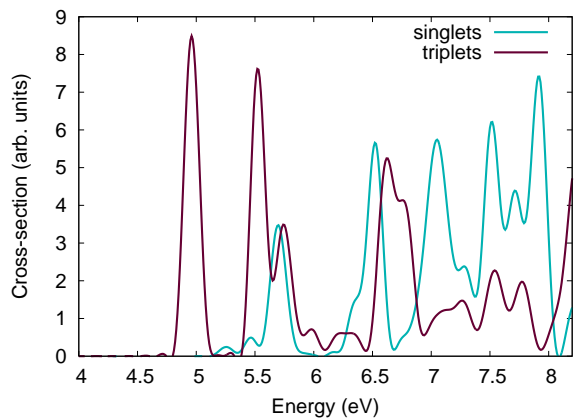,width=0.45\textwidth}
  \caption{(Color online) Si$_8$H$_{18}$ averaged absorption
    cross-section for singlets and triplets.  Again, The curves
    obtained with and without the use of symmetry completely overlap.
    \label{fig:si8h18_response}}
\end{figure}

To represent the wave-functions in real space we used a uniform grid
with a spacing of 0.195\,\AA\ and a box composed by spheres of radius
5\,\AA\ around every atom.  In order to propagate the Kohn-Sham
orbitals we employed state of the art
algorithms\cite{propagators}. A time step of 0.0048\,fs
assured the stability of the time propagation, and a total propagation
time of 19.35\,fs allowed a resolution of about 0.1\,eV in the
resulting spectrum.

The results obtained are summarized in Fig.~\ref{fig:tpg_response} and
Fig.~\ref{fig:si8h18_response} where we plot the averaged absorption
cross-section (trace of the tensor) for singlets and triplets.

\section{Conclusions}
\label{section:conclusions}

In summary, we presented a scheme that can considerably reduce the
number of calculations necessary to obtain the full polarizability
tensor by using the symmetries of the system. These can be spatial
symmetries (like planes of inversion or symmetry axis, e.g.) or they
can lie in spin space (if the ground-state is spin
saturated). Finally, the scheme is trivial to implement, and can be
easily extended to different symmetries and different responses--
i.e. higher multipole responses, and higher order
hyperpolarizabilities --, so we expect that its usefulness to surpass
the present context.

\begin{acknowledgments}
We thank Claudia Cardoso for providing the molecular geometry of
protonated triphenylguanidine. This work has being funded by the EC
Network of Excellence NANOQUANTA (ref. NMP4-CT-2004-500198), the
Spanish Ministry of Education (grant FIS2004-05035-C03-03), the SANES
(ref. NMP4-CT-2006-017310), DNA-NANODEVICES (ref. IST-2006-029192),
and NANO-ERA Chemistry projects, the University of the Basque Country
EHU/UPV (SGI/IZO-SGIker UPV/EHU Arina), and the Basque Country. We
thankfully acknowledge the computer resources, technical expertise and
assistance provided by the Barcelona Supercomputing Center - Centro
Nacional de Supercomputaci\'{o}n and by the Laboratory for Advanced
Computation of the University of Coimbra. A.C. acknowledges support by
the Deutsche Forschungsgemeinschaft in SFB 658 and MJTO acknowledges
financial support from the portuguese FCT (contract
\#SFRH/BD/12712/2003).

\end{acknowledgments}

\end{document}